\documentclass[aps,prd,showpacs,twocolumn,superscriptaddress,nofootinbib,preprintnumbers]{revtex4-1} 
\newcommand{\eV}{\,\text{eV}}

\newcommand{\eff}{{\rm eff}}

\newcommand{\wa}{w_0w_a{\rm CDM}}
\newcommand{\ld}{\Lambda{\rm CDM}}

\usepackage{float}

\usepackage{amssymb, amsmath, bm, dcolumn, epsf, graphicx, latexsym, slashed} 
\usepackage[utf8]{inputenc}
\usepackage{multirow}

\usepackage{color}

\def\be{\begin{equation}}
\def\ee{\end{equation}}
\def\bea{\begin{eqnarray}}
\def\eea{\end{eqnarray}}

\usepackage{hyperref}
\usepackage{comment}
\usepackage{booktabs}

\usepackage[normalem]{ulem}

\begin{document}


\title{
Modified gravity interpretation of the evolving dark energy in light of DESI data
}

\author{Anton Chudaykin}\email{anton.chudaykin@unige.ch}
\affiliation{D\'epartement de Physique Th\'eorique and Center for Astroparticle Physics,\\
Universit\'e de Gen\`eve, 24 quai Ernest  Ansermet, 1211 Gen\`eve 4, Switzerland}
\author{Martin Kunz}\email{martin.kunz@unige.ch}
\affiliation{D\'epartement de Physique Th\'eorique and Center for Astroparticle Physics,\\
Universit\'e de Gen\`eve, 24 quai Ernest  Ansermet, 1211 Gen\`eve 4, Switzerland}

\begin{abstract} 
The Dark Energy Spectroscopic Instrument (DESI) collaboration has recently released measurements of baryon acoustic oscillation (BAO)
from the first year of observations.
A joint analysis of DESI BAO, CMB, and SN Ia probes indicates a preference for time-evolving dark energy. 
We evaluate the robustness of this preference by replacing the DESI distance measurements at $z<0.8$ with the SDSS BAO measurements in a similar redshift range.
Assuming the $\wa$ model, we find an evolution of the dark energy equation of state parameters consistent with $\Lambda$CDM.
Our analysis of $\chi^2$ statistics across various BAO datasets shows that DESI's preference for evolving dark energy is primarily driven by the two LRG samples at $z_\eff=0.51$ and $z_\eff=0.71$, with the latter having the most significant impact.

Taking this preference seriously, we study a general Horndeski scalar-tensor theory, which provides a physical mechanism to safely cross the phantom divide, $w=-1$.
Utilizing the Effective Field Theory of dark energy and adopting the $\wa$ background cosmological model, we derive constraints on the parameters $w_0=-0.856\pm0.062$ and $w_a=-0.53_{-0.26}^{+0.28}$ at $68\%$ CL. from Planck CMB, Planck and ACT CMB lensing, DESI BAO, and Pantheon+ datasets, showing good consistency with the standard $\wa$ model.
The modified gravity model gives results discrepant with $\Lambda$CDM at the $2.4\sigma$ level, while for $\wa$ it is at $2.5\sigma$, based on the best-fit $\chi^2$ values.
We conclude that modified gravity offers a viable physical explanation for DESI's preference for evolving dark energy.
\end{abstract}

\maketitle

\section{Introduction}
\label{sec:intro}

Recent measurements of baryon acoustic oscillations (BAO) from the first year of observations from the Dark Energy Spectroscopic Instrument (DESI) find hints towards a time-evolving equation of state of dark energy~\cite{DESI:2024mwx}.
The BAO measurements have been obtained in seven redshift bins using galaxy, quasar, and Lyman-$\alpha$ forest tracers across $0.1 < z < 4.2$
They are expressed in terms of the transverse comoving distance $D_M/r_d$, the Hubble distance, $D_M/r_d$, or their combination, the angle-averaged distance $D_V/r_d$, normalized to the comoving sound horizon at the end of baryon drag epoch.
While the DESI BAO data alone are consistent with the standard $\Lambda$CDM model, the combination with external datasets yield results discrepant with the concordance cosmological scenario.

Allowing for a time-varying dark energy equation of state, parameterized by $w(a)=w_0+w_a(1-a)$, combinations of DESI with cosmic microwave background (CMB) or type Ia supernovae measurements individually favours $w_0>-1$ and $w_a<0$.
This preference is $2.6\sigma$ for the DESI+CMB data, and increases when adding supernova measurements.
A joint analysis of DESI, CMB, and SN Ia probes gives results discrepant with the $\Lambda$CDM model at the $2.5\sigma$, $3.5\sigma$ and $3.9\sigma$ levels when using Pantheon+~\cite{Scolnic:2021amr}, Union3~\cite{Rubin:2023ovl}, or Dark Energy Survey (DES) Y5~\cite{DES:2024tys} supernova datasets, respectively. 

The DESI collaboration reported systematically larger observed BAO scales than the prediction of Planck 2018 $\Lambda$CDM cosmology at $z<0.8$, and therefore lower distances.
For LRG1 at $z_\eff=0.51$, the DESI data prefers a mildly larger line-of-sight BAO scale compared to the transverse scale, leading to a smaller $D_H/D_M$ compared than the Planck $\Lambda$CDM prediction. 
For LRG2 at $z_\eff=0.71$, the discrepancy is sourced from the transverse BAO scale, implying a smaller $D_M/r_d$ and $D_V/r_d$ at this redshift.
Accordingly, DESI reported an approximate $3\sigma$ discrepancy between the DESI BAO measurement at $z_\eff=0.71$ and the SDSS measurement at $z_\eff=0.7$~\cite{DESI:2024mwx}.~\footnote{The SDSS catalog at $z_\eff=0.7$ is a composite dataset formed from  the BOSS CMASS galaxy sample and the eBOSS LRG sample, for details see~\cite{eBOSS:2020yzd,eBOSS:2020mzp}.}
Given the consistency of the DESI measurements with SDSS at $z>0.8$~\cite{DESI:2024uvr}, the difference between the two BAO datasets originates solely from $z<0.8$.
This observation suggests an important consistency check of the DESI results, which involves replacing the low-redshift DESI data at $z<0.8$ with the SDSS BAO measurements.

The DESI collaboration preformed a consistency test by replacing the data at $z<0.6$ with SDSS while using a combined ``DESI+SDSS'' BAO dataset at $z>0.6$.
This test showed a good agreement with the baseline analysis that utilized the entire DESI data.
Notably, the constraints in the $w_0$-$w_a$ plane shifted closer to the $\Lambda$CDM expectation, and the significance of the tension with $\Lambda$CDM marginally decreased to $2.1\sigma$ with the Pantheon+ data~\cite{DESI:2024mwx}.
Crucially, this test retained the $z_\eff=0.71$ DESI point, which shows the greatest discrepancy with both the Planck $\ld$ expansion history and the SDSS distance measurement at $z_\eff=0.71$.
Given the previous studies~\cite{Wang:2024pui,DESI:2024uvr} suggesting a significant role of the LRG2 in driving the DESI preference for the $w_0w_a$-model relative to $\ld$, replacing this point with the SDSS data is capable of reducing the evidence for time-evolving dark energy as we will show in Section \ref{sec:res1} below.

One particular aspect of the preference of the DESI+CMB+SN data for an evolving dark energy equation of state is that the preferred models cross the `phantom divide', $w=-1$, a scenario supported by some dark energy models~\cite{DESI:2024kob} and model-agnostic expansion history reconstructions~\cite{DESI:2024aqx,Yang:2024kdo}. 
To some extent this is driven by the CMB data, which fixes the distance to last scattering to be close to the $\Lambda$CDM expectation, which requires to compensate a positive $1+w$ at low redshift with a negative value at higher redshift~\cite{Linder:2007ka}. This adds an additional challenge when trying to interpret the DESI results in terms of physical models.
The $w_0w_a$-model offers a phenomenological description of the equation of state of dark energy at the background level, but it fails to provide a satisfactory description for dark energy perturbations. The dark energy perturbations are usually taken to be those of an canonical minimally-coupled scalar field component, but this model
lacks a physical mechanism to safely cross the phantom divide.
To avoid singularities in the fluid equation of motion,
one commonly employs the parameterized post-Friedman approach to metric perturbations~\cite{Fang:2008sn}. 
Although this framework is extensively used in the literature, it provides only an effective prescription of the perturbation evolution and is not based on a full physical model.
Hence, a physical model for dark energy perturbations is necessary.

The phantom crossing in the dark energy equation of state can be naturally realized in models involving multiple scalar fields, alternative models of gravity with Kinetic Gravity Braiding, or with non-minimal couplings. For the purposes of this analysis, we consider a Horndeski theory of gravity with an additional scalar degree of freedom, which features at most second-order derivatives in the equations of motion.
This class of theories is quite general and encompasses a large groups of popular dark energy and modified gravity models,
including $f(R)$ gravity, Brans-Dicke theory, quintessence, k-essence, Galileons, and many other more exotic possibilities and combinations, some of which can cross the phantom divide self-consistently.
Given the extensive landscape of modified gravity theories, it is beneficial to use a flexible framework that allows for a model-independent description of the effects of modified gravity.

An efficient and systematic description of the linear perturbations in a general Horndeski theory is provided by the Effective Field Theory (EFT) of dark energy.
In this approach, all the possible modifications of the Einstein-Hilbert action, which govern the evolution of linear perturbations, are parameterized by a set of fixed operators that respect the symmetries of the cosmological background in unitary gauge.  
The contribution of each operator is controlled by EFT coefficients, which are functions of time only and are often referred to as “$\alpha$-functions”.
On the EFT grounds, the $\alpha$-functions are arbitrary functions of time.
However, current astrophysical data is not sufficiently constraining for general functions.
Consequently, a common approach is to assume specific parametrizations for the $\alpha$-functions, thereby reducing their time-dependence to a set of well-defined parameters.
Constraints on these parameters serve as a test for deviations from General Relativity in a largely model-independent manner.
Another advantage of the EFT-based approach is that the background expansion is entirely independent of the dynamics of linear perturbations, offering a more flexible framework for testing non-GR effects at the perturbative level.

In this work, we investigate the preference for a dynamical dark energy from the DESI BAO DR1 data in combination with external datasets including Planck CMB, Planck+ACT CMB lensing, DESI and SDSS BAO measurements, and the Pantheon+ supernova sample.
First, we identify the source of this preference in the DESI data assuming the $\wa$ model. 
Second, we evaluate the robustness of these results by replacing the low-redshift DESI BAO data with the SDSS measurements.
Third, we evaluate parameter estimates within a general Horndeski theory, allowing for a reasonable degree of freedom in the dynamics of linear perturbations within the EFT of dark energy approach.
For the sake of comparison with the $\wa$ scenario, we set the background cosmological model to $\wa$.

Our paper is structured as follows. 
We describe the analysis procedure and data in Sec.~\ref{sec:data} 
Our final results are presented in Sec.~\ref{sec:res}.
We draw conclusions in Sec.~\ref{sec:conc}.
Additional parameter constraints are given in Appendix~\ref{app:full}.

\section{Analysis procedure}
\label{sec:data}

We investigate dark energy dynamics with the $\wa$ model and the general scalar-tensor Horndeski theory.
In $\wa$, we utilize the standard parametrization for the dark energy equation of state, given by $w(a)=w_0+w_a(1-a)$.
For Horndeski gravity, we adopt a model-independent parametrization based on the Effective Field Theory (EFT) of dark energy.
Hereafter, we will refer to this modified gravity model as ``MG''. 

\subsection{EFT of dark energy}
\label{sec:data1}

In the EFT-based approach to modified gravity, the dynamics of linear perturbations associated with a scalar degree of freedom is parameterized by four arbitrary functions of time, termed $\alpha$-functions. 
These include the kineticity parameter $\alpha_K$, which modulates the kinetic term for scalar field perturbations, the braiding parameter $\alpha_B$, which controls a mixing between the kinetic terms of the metric and the scalar field, the running of the effective Planck mass $\alpha_M$, and the tensor speed excess $\alpha_T$, which captures the difference between the propagation speed of gravitational waves and the speed of light.~\footnote{We set the effective Planck mass at early times to be $M_{\rm Pl}$, since we focus on late-time modifications of gravity and do not aim to simultaneously constrain early-time modifications. The initial value of the effective Planck mass is tightly constrained by CMB data~\cite{Bellini:2015xja}.}

In order to extract meaningful cosmological constraints within the EFT-based framework, it is necessary to specify a parametrization for the $\alpha$-functions.
We set the time-dependence of the $\alpha$-functions to match the density fraction of dark energy, $\alpha_i(a)=c_i\Omega_{DE}(a)$, where $c_i$ is a free constant parameter. 
This parametrization ensures that $\alpha_i\to0$ at early time, consistent with the expectation that any modifications of gravity are suppressed during epochs when dark energy does not significantly contribute to the energy density of the Universe.
Additionally, as mentioned, we assume a $w_0w_a$CDM background cosmology.

The general form of the $\alpha_i(a)$ and the background expansion $H(a)$ within the EFT-based framework can lead to ill-defined theories.
To prevent pathological behavior, we enforce several stability conditions that guarantee the absence of ghost and gradient instabilities, as implemented in \texttt{hi\_class}~\cite{Zumalacarregui:2016pph}.
Tachyon instabilities are considered less pathological, since they appear in the low-$k$ limit, and come under control when the mode enters the sound horizon (see~\cite{Frusciante:2018vht} and references therein). 
Therefore, we adopt a conservative approach and do not impose the no-tachyon condition, allowing the data to exclude such scenarios.

\subsection{Methodology}
\label{sec:data2}

Parameter estimates in this paper are obtained with the publicly available code \texttt{hi\_class}~\cite{Zumalacarregui:2016pph,Bellini:2019syt}, interfaced with the \texttt{Montepython} Monte Carlo sampler.
\texttt{hi\_class} is an extension of the Boltzmann code CLASS~\cite{Blas:2011rf} and calculates the evolution of linear cosmological perturbations for general Horndeski scalar-tensor gravity theories.
Modified gravity models can be specified either within the Effective Field Theory of Dark Energy approach or through the full covariant action.
In this work, we utilize the EFT-based parameterized approach.

We perform Markov chain Monte Carlo (MCMC) analyses to sample the posterior distributions using the Metropolis-Hastings algorithm. 
We adopt a Gelman-Rubin~\cite{Gelman:1992zz} convergence criterion $R-1<0.02$ (unless otherwise stated).
The plots and marginalized constraints are generated with the publicly available \texttt{getdist} package.~\footnote{\href{https://getdist.readthedocs.io/en/latest/}{https://getdist.readthedocs.io/en/latest/}}

In the minimal $\Lambda$CDM model we vary 6 standard cosmological parameters within broad uniform priors: $\omega_{cdm}=\Omega_ch^2$, $\omega_b=\Omega_bh^2$, $H_0$, $\ln(10^{10}A_s)$, $n_s$ and $\tau$. 
In the $w_0w_a$CDM scenario we sample two additional parameters, $(w_0,w_a)$, which parameterize a time-varying equation of state for dark energy.
We adopt the uniform priors on the dark energy parameters, $w_0\in[-3,1]$ and $w_a\in[-3,2]$, as specified in~\cite{DESI:2024mwx}.
Since we explore the parameter space that allows the dark energy equation of state to cross the $w=-1$ boundary, we use the parameterized post-Friedman approach~\cite{Fang:2008sn} to compute dark energy perturbations in the $w_0w_a$CDM scenario. 
Additionally, we impose the condition $w_0+w_a<0$, ensuring a matter domination stage in the past.
In Horndeski theory, we additionally vary two EFT amplitudes, $c_B$ and $c_M$, within broad uniform priors while fixing $c_K=1$.
We also set $c_T=0$ to address the tight constraint on the speed of propagation of gravitational waves.
In all cases, we assume a spatially flat universe with two massless and one massive neutrino species with $m_\nu=0.06\eV$. 
In the $\Lambda$CDM and $\wa$ models, we use the Halofit module to compute the non-linear matter power spectrum.
Since the Halofit fitting formula was not calibrated for modified gravity scenarios, we conservatively adopt the linear prediction within the general Horndeski theory.~\footnote{Employing the Halofit module in the MG scenario does not significantly affect the posterior distribution. In particular, the shifts in cosmological parameters w.r.t the baseline analysis do not exceed $0.5\sigma$ in terms of statistical uncertainty, with the modified gravity parameters shifting by less than $0.2\sigma$.}

In our analyses we compute the best-fit predictions as follows. 
We perform an additional MCMC by multiplying the log-posteriors by a factor 100
and rescaling the covariance of the proposal distribution by a factor of 0.01.
This approach enhances the efficiency of sampling in the high probability region, thereby improving the precision of the best-fit model recovery.
We assess the uncertainty in the calculation of the minimum $\chi^2$ to be $0.2$ by performing 10 independent minimizations.
This level of accuracy is sufficient to reach robust conclusions.

\subsection{Data}
\label{sec:data3}
We consider the following datasets in our MCMC analyses:

\subsubsection{CMB}

We employ the official Planck high-$\ell$ \texttt{plik} TT,TE,EE likelihood, together with low-$\ell$ TT \texttt{Commander} and EE \texttt{SimAll} likelihoods~\cite{Planck:2018vyg}. 
In addition to the primary CMB anisotropy, we include the measurement of the lensing potential power spectrum, $C^{\phi\phi}_\ell$, from the combination of \texttt{NPIPE} PR4 Planck CMB maps~\cite{Carron:2022eyg} and the Data Release 6 of the Atacama Cosmology Telescope (ACT)~\cite{ACT:2023kun}.~\footnote{The likelihood is publicly available at \href{https://github.com/ACTCollaboration/act\_dr6\_lenslike}{https://github.com/ACTCollaboration/act\_dr6\_lenslike}; we use \texttt{actplanck\_baseline} option.} We will refer to the CMB data simply as ``CMB''.

Recently, a new Planck data release, PR4, has been made available, involving a reprocessing of the data from both the LFI and HFI channels with the latest and most accurate pipeline, \texttt{NPIPE}~\cite{Planck:2020olo}. This release features lower levels of noise and systematics, as well as slightly more data. In our analysis, we use the official PR3 \texttt{plik} likelihood for the following reasons.
First, shifts in parameter estimates inferred from the PR3 and PR4 Planck likelihoods become non-significant after combining with CMB lensing likelihoods from Planck and ACT. 
Second, we retain \texttt{plik} as our baseline to closely align with the DESI analysis~\cite{DESI:2024mwx}.

\subsubsection{BAO}

In our baseline analysis we utilize the DESI DR1 BAO data as implemented in the official DESI likelihood.~\footnote{The DESI likelihood is publicly available at \href{https://github.com/cosmodesi/desilike}{https://github.com/cosmodesi/desilike}}
This dataset includes the BGS sample in the redshift range $0.1<z<0.4$, LRG1 and LRG2 samples in $0.4<z<0.6$ and $0.6<z<0.8$, combined LRG3+ELG1 sample in $0.8<z<1.1$, ELG2 sample in $1.1<z<1.6$, quasar sample in $0.8<z<2.1$ and Lyman-$\alpha$ Forest sample spanning $1.77<z<4.16$.
We will refer to this BAO data as ``DESI BAO''.

To explore the effect of low-redshift BAO measurements, 
we construct a combined dataset which comprises SDSS BAO data in the low-redshift region $z<0.8$ and DESI BAO data in high-redshift region $z>0.8$.
In particular, we adopt the SDSS BAO distance measurements extracted from DR7 Main Galaxy Sample (MGS)~\cite{Ross:2014qpa} in $0.07<z<0.2$, BOSS DR12 galaxy samples in $0.2<z<0.5$ and $0.4<z<0.6$~\cite{BOSS:2016wmc}, eBOSS DR16 galaxy sample in $0.6<z<1.0$~\cite{eBOSS:2020yzd} in place of the DESI BGS, LRG1 and LRG2 points.
We use the label ``DESI$^\star\!$+SDSS'' to refer to this composite dataset.

We summarize the BAO measurements used in this work in Tab.~\ref{tab:BAO}.
\begin{table}[!t]
    \renewcommand{\arraystretch}{1.2}
    \centering
    \caption{BAO measurements used in our analysis along with their tags, effective redshift, and number of data points. The upper group contains the DESI BAO data~\cite{DESI:2024mwx}, while the bottom group refers to the SDSS BAO measurements~\cite{eBOSS:2020yzd}.}
    $\,$\\ 
    \begin{tabular}{c|c|c} \hline  \hline 
         Data & 
         $z_\eff$ &
         $N_{\rm data}$ \\
         \hline  
         BGS & 
         $0.30$ & 
         $1$ \\ 
         LRG1 & 
         $0.51$ & 
         $2$ \\
         LRG2 & 
         $0.71$ & 
         $2$ \\
         LRG3+ELG1 & 
         $0.93$ & 
         $2$ \\
         ELG2 & 
         $1.32$ & 
         $2$ \\
         QSO & 
         $1.49$ & 
         $1$ \\
         Lya & 
         $2.33$ & 
         $2$ \\
         \hline 
         MGS & 
         $0.15$ & 
         $1$ \\
         BOSS DR12 & 
         $0.38\,\text{and}\,0.51$ & 
         $4$ \\
         eBOSS DR16 & 
         $0.7$ & 
         $2$ \\
         \hline \hline 
    \end{tabular}
\label{tab:BAO}
\end{table}

\subsubsection{SN}

We utilize supernova data from the Pantheon+ sample~\cite{Scolnic:2021amr}, which consists of 1550 spectroscopically confirmed SNe I in the redshift range $0.001<z<2.26$.
We do not apply any external calibration from astrophysical measurements and allow a supernova absolute magnitude to vary freely.
We adopt the \texttt{Pantheon\_Plus} likelihood in \texttt{MontePython} and recall this dataset as ``SN'' in what follows.

\section{Results and Discussions}
\label{sec:res}

\subsection{$w_0w_a$CDM model}
\label{sec:res1}

In this section, we explore a time-varying dark energy equation of state, $w(a)=w_0+w_a(1-a)$, referred to as $w_0w_a$CDM.

\begin{figure}[!t]
\includegraphics[keepaspectratio,width=1\columnwidth]{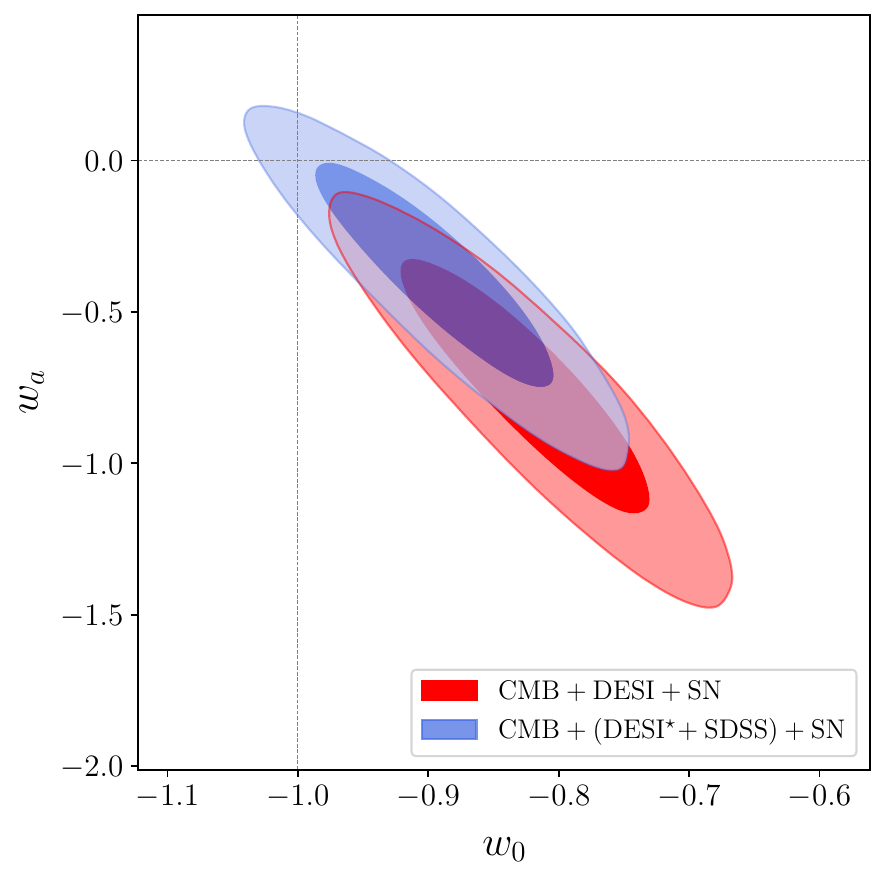}
\caption{
68\% and 95\% marginalized posterior constraints in the $w_0-w_a$ plane for the $w_0w_a$CDM model from $\rm CMB+DESI+SN$ (red) and $\rm CMB+(DESI^\star\!+SDSS)+SN$ (blue) datasets.}
\label{fig:w0wa} 
\end{figure}

Fig.~\ref{fig:w0wa} presents the marginalized constraints in the $w_0-w_a$ plane from different data.
We start by analyzing cosmological constraints from the $\rm CMB+DESI+SN$ combination utilized in the official DESI analysis.
The resulting $68\%$ credible-interval constraints are
\be
\left.
\begin{aligned}
&w_0=-0.821\pm0.065,\\
&w_a=-0.76_{-0.26}^{+0.31},
\end{aligned}
\ \right\} \quad \mbox{$\rm CMB\!+\!DESI\!+\!SN$}
\ee
Our parameter estimates align closely with the results of the official DESI analysis.
The minimum $\Delta\chi^2$ value between the $w_0w_a$CDM and $\Lambda$CDM models is $-8.8$, 
indicating a preference for an evolving dark energy equation of state at the $2.4\sigma$ level.~\footnote{
The likelihood-ratio test applied here evaluates how likely the observed outcome was under the null hypothesis compared to the alternative.
This does not assign probabilities to the models themselves. We interpret the resulting p-value in terms of a $\sigma$-interval to provide a more intuitive result.
}

Next, we examine the robustness of the preference for an evolving dark energy using an alternative BAO configuration.
Specifically, we replace the low-redshift DESI points at $z<0.8$ with the SDSS BAO measurements, which provide more precise distance measurements in this redshift range.
The resulting constraints on dark energy parameters are
\be
\left.
\begin{aligned}
&w_0=-0.891\pm0.062,\\
&w_a=-0.39_{-0.23}^{+0.27},
\end{aligned}
\ \right\} \, \mbox{\parbox{0.2\textwidth}{\centering$\rm CMB\!+\!(DESI^\star\!+\!SDSS)$\\$\rm +SN$}}
\ee
The dark energy equation of state today is now consistent with $w=-1$ at the $1.8\sigma$ level.
The inclusion of the low-$z$ SDSS data also shifts the $w_a$ posterior towards zero, making it more consistent with the cosmological constant. 
The minimum $\Delta\chi^2$ value between the $w_0w_a$CDM and $\Lambda$CDM models is $-3.0$ for 2 additional degrees of freedom.
Given the small $\Delta\chi^2$, there is no significant evidence for an evolving dark energy equation of state when using the DESI$^\star\!$+SDSS setup.

\begin{figure}[!t]
\includegraphics[keepaspectratio,width=1\columnwidth]{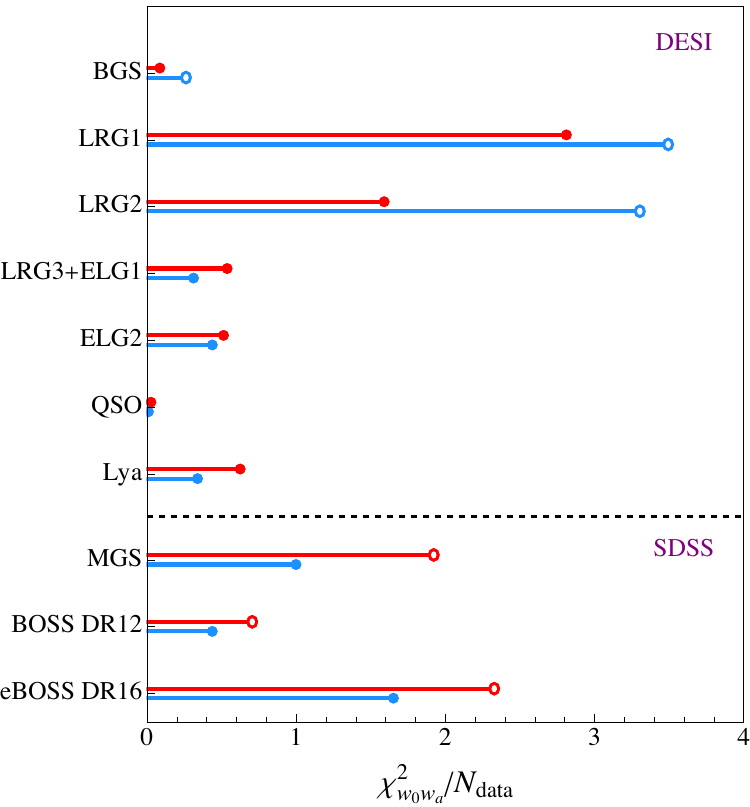}
\caption{
$\chi^2/N_{\rm data}$ values for the individual BAO measurements computed in the best-fit $w_0w_a$CDM model to the $\rm CMB+DESI+SN$ (red) and $\rm CMB+(DESI^\star\!+SDSS)+SN$ (blue) data. $N_{\rm data}$ represents the number of data points in a given experiment, as provided in Tab.~\ref{tab:BAO}.
Results for the BAO data not used in the fitting procedure are displayed with open markers.
}
\label{fig:chi2} 
\end{figure}

To understand the source of the differences in parameter constraints, it is necessary to look at the breakdown of the minimum $\chi^2$ across various BAO datasets.
Fig.~\ref{fig:chi2} shows the $\chi^2/N_{\rm data}$ values for the individual BAO measurements in the best-fit $w_0w_a$CDM models.
For the full DESI data analysis, the largest contributions to the total $\chi^2$ come from LRG1 and LRG2 points, with values of $\chi^2=5.6$ and $3.2$, respectively.
When using the $\rm (DESI^\star\!+SDSS)$ data, we found a reasonable fit to each individual BAO measurements involved in the fitting procedure.
Interestingly, this noticeably degrades the fit to the DESI measurements that were not included in the fitting procedure.

The first important observation is that the fit to LRG1 does not significantly improve after taking into account the entire DESI data. 
This suggests that the DESI BAO measurement at $z_\eff=0.51$ is too extreme to be fitted with a smooth $(w_0,w_a)$ parameterization, and thus, the addition of the LRG1 point has little impact on the constraints in the $\wa$ scenario.
This outcome is consistent with previous studies~\cite{Wang:2024rjd,Wang:2024pui}, which reported very consistent constraints after excluding the LRG1 point.
Therefore, although the LRG1 degrades the fit, it cannot explain the DESI preference for an evolving dark energy~\cite{DESI:2024mwx}.

Another important observation relates to the LRG2 data point. 
When this measurement is excluded, the $\wa$ model shows a poor fit to the LRG2, with a minimum $\chi^2=6.6$.
The fit quality improves significantly to $\chi^2=3.2$, when the entire DESI dataset is considered.
This behaviour can be traced to the discrepancy between the LRG2 and eBOSS DR16 results.
The level of disagreement between these datasets is close to $3\sigma$, mainly for $D_M/r_d$ and $D_V/r_d$~\cite{DESI:2024mwx}.
Given that the eBOSS DR16 is fully consistent with the Planck best-fit $\Lambda$CDM cosmology~\cite{eBOSS:2020yzd}, replacing the DESI BAO measurement at $z_\eff=0.71$ with the SDSS measurement at $z_\eff=0.70$ shifts the parameter constraints in the $(w_0,w_a)$ plane towards $\ld$. 


To further explore this preference, we calculate the $\Delta \chi^2_{w_0w_a}$, the difference between the best-fit $\wa$ and $\ld$ models to the CMB+DESI+SN data.
We found the cumulative value across all DESI BAO measurements to be $\Delta \chi^2_{w_0w_a}({\rm DESI})=-3.8$. This difference is largely driven by an improvement of the fit to the LRG2 data point, with $\Delta \chi^2_{w_0w_a}({\rm LRG2})=-3.3$.
The second largest contribution comes from LRG1, yielding $\Delta \chi^2_{w_0w_a}({\rm LRG1})=-1.8$.~\footnote{Note that $\ld$ provides a better fit to the LRG3+ELG1, ELG2 and Lya data points, with the overall contribution $\Delta \chi^2_{w_0w_a}=+1.4$ for these datasets.}
Our finding align with those of~\cite{Wang:2024pui}, showing that the LRG1 and LRG2 mainly contribute into the preference for an evolving dark energy.

The difference in the parameter constraints when altering the low-redshift BAO measurements raises concerns about the robustness of the preference for time-evolving dark energy in the DESI data. 
Future DESI DR3 and Euclid \cite{Euclid:2024yrr} BAO data will help to clarify whether this evidence necessitates new physics in the dark sector or is merely a statistical fluctuation. 
For now, we will assume that this evidence is real and utilize the entire set of DESI distance measurements in what follows.

\subsection{MG model}
\label{sec:res2}

We now explore a general Horndeski scalar-tensor theory that in principle allows 
the dark energy equation of state to evolve across the phantom divide, $w=-1$.
First of all, we found that our results are not sensitive to the kineticity parameter $c_K$.
This is expected, since $c_K$ 
has minimal impact on sub-horizon physics~\cite{Bellini:2014fua,Noller:2018wyv}.
For definiteness, we fix $c_K=1$ in our analysis.

\begin{figure*}[!t]
\includegraphics[keepaspectratio,width=1.2\columnwidth]{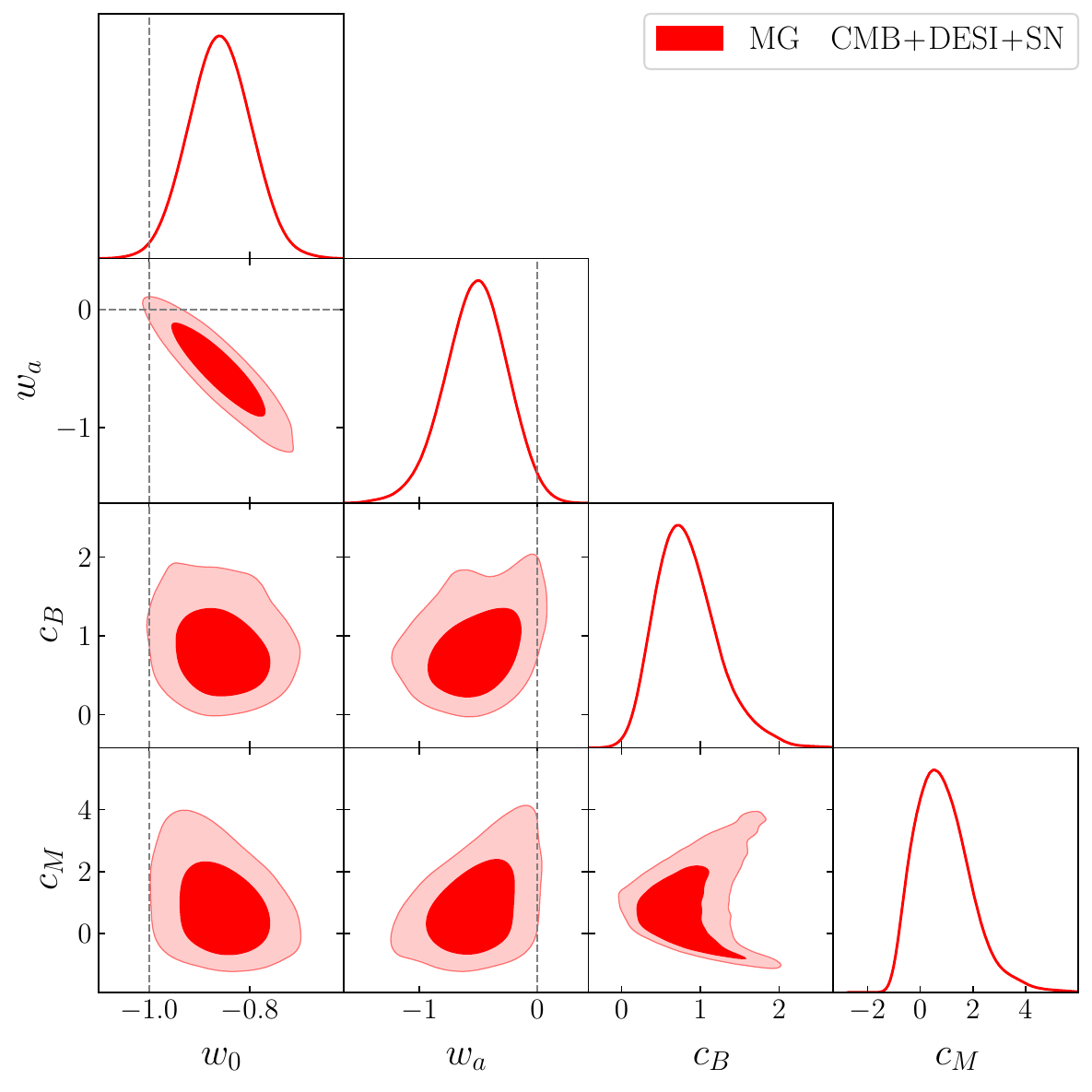}
\caption{
68\% and 95\% marginalized constraints on $w_0$, $w_a$, $c_B$ and $c_M$ from the CMB+DESI+SN data in the MG model.}
\label{fig:MG} 
\end{figure*}

Fig.~\ref{fig:MG} shows the posterior distributions in the $w_0-w_a-c_B-c_M$ space for the MG model.
The one-dimensional parameter constraints are
\be
\label{eq:kgb}
\left.
\begin{aligned}
&w_0=-0.856\pm0.062,\\
&w_a=-0.53_{-0.26}^{+0.28},\\
&c_B=0.82_{-0.46}^{+0.31},\\
&c_M=0.88_{-1.28}^{+0.75}.
\end{aligned}
\ \right\} \quad \mbox{$\rm CMB\!+\!DESI\!+\!SN$}
\ee
Full parameter estimates from various datasets are provided in App.~\ref{app:full}.

The MG model offers a satisfactory physical description of time-evolving dark energy, allowing for the crossing of the phantom divide, $w = -1$.
In particular, the $w_0$ posterior deviates from $-1$ at the $2.4\sigma$ level.
The minimum $\Delta\chi^2$ value between the MG model and the concordance cosmology is $-12.2$ for 4 additional degrees of freedom.
This reflects a slightly lower preference for evolving dark energy, when compared to the $\wa$ scenario, at the $2.4\sigma$ level relative to $\ld$.

Constraints on the MG parameters are primarily driven by the late integrated Sachs-Wolfe (ISW) effect.
The ISW effect especially excludes cosmologies with large $c_B$ and $c_M$ values, as it leads to a large excess of power in the TT CMB power spectrum on large scales~\cite{Zumalacarregui:2016pph}.
For not very large $c_i$, the effect of modified gravity conspires to prevent large power in the TT CMB power spectrum, resulting in the particular degeneracy directions in the $c_M$-$c_B$ plane~\cite{Noller:2018wyv}.
The data shows a mild preference for positive values of the braiding parameter, $c_B>0.11$ at $95\%$ CL., in agreement with the previous study~\cite{Noller:2018wyv}. 
The lower boundary for $c_M$ is driven by the onset of gradient instabilities.

While $c_M=0$ provides a good fit to the data, non-minimal coupling play an important role in the MG constraints.~\footnote{Ref.~\cite{Wolf:2024stt} highlights the importance of non-minimal gravity coupling in explaining DESI's preference for evolving dark energy.}
For the scalar minimally coupled to gravity ($c_M=0$), the region $c_B<0.4$ is excluded due to the presence of gradient instabilities. 
A non-zero $c_M$ shifts the gradient stability condition, making cosmologies with small $c_B$ viable.
The region of small $c_B$ is then disfavoured by the data, as it implies too large $c_M$, which leads to large power in the $C_\ell^{\rm TT}$ via the late-time ISW effect.
Thus, non-minimal coupling provides more flexibility, enabling the data to explore the full parameter space in the MG scenario.

It is important to note that measurements of the late-time fluctuation amplitude significantly improves the MG constraints by breaking the degeneracy directions in the $c_M$-$c_B$ plane, as discussed in~\cite{Noller:2018wyv}.
In this work, our primary goal is to demonstrate a proof-of-principle theory that accommodates the crossing of the phantom divide, evidenced by the CMB+DESI+SN data.
To streamline the analysis, we focus here on the distance measurements.
We leave a detailed analysis of the MG scenario with all available data to future work.

\subsection{Comparison And Discussion}
\label{sec:res3}

Here, we compare the $\wa$ and MG models for the CMB+DESI+SN dataset.
Results for other datasets are available in App.~\ref{app:full}.

\begin{figure}[!t]
\includegraphics[keepaspectratio,width=1\columnwidth]{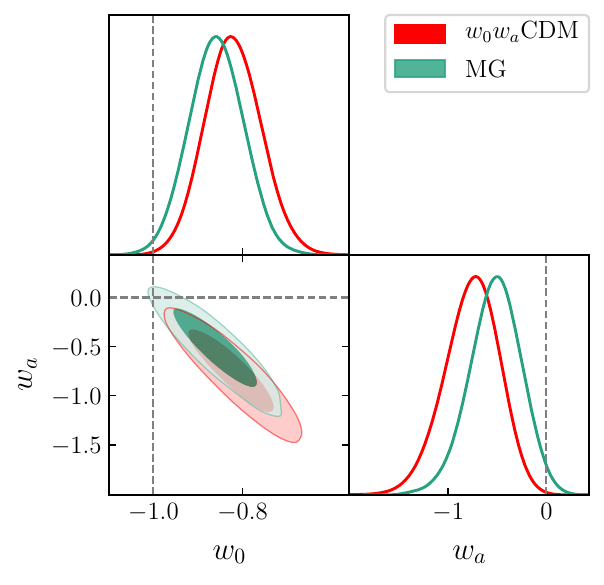}
\caption{
68\% and 95\% marginalized constraints on $w_0$ and $w_a$ from the CMB+DESI+SN data in $\wa$ (red) and MG (green) models.}
\label{fig:comp} 
\end{figure}

Fig.~\ref{fig:comp} shows the posterior distributions in the $w_0$-$w_a$ plane for the $\wa$ and MG models.

The MG scenario shows a slightly weaker preference for evolving dark energy,
as evidenced by the $w_0$ and $w_a$ posteriors shifting towards $\Lambda$CDM. 
This behaviour is mainly attributed to the CMB degeneracies between the dark energy parameters and $c_B$ shown in Fig.~\ref{fig:MG}.
However, the results from the MG and $\wa$ models are fully compatible.

We also compare the goodness-of-fit for different datasets between the $\wa$ and MG models relative to $\ld$.
Tab.~\ref{tab:chi2} presents the $\Delta \chi^2$ for various datasets in the CMB+DESI+SN analysis.
\begin{table}[!t]
    \renewcommand{\arraystretch}{1.2}
    \centering
    \caption{$\Delta \chi^2_{\rm model} \equiv \chi^2_{\rm model}-\chi^2_{\Lambda\rm CDM}$ values for the different best-fit models optimized to the $\rm CMB+DESI+SN$ data.
    The minimum $\chi^2$ values were obtained using the Halofit non-linear matter power spectrum, see discussion in the text.}
    $\,$\\ 
    \begin{tabular}{c|c|c} \hline  \hline 
         Data & 
         $\Delta \chi^2_{w_0w_a}$ &
         $\Delta \chi^2_{\rm MG}$ \\
         \hline  
         CMB & 
         $-2.9$ & 
         $-5.9$ \\
         DESI & 
         $-3.8$ & 
         $-4.0$ \\
         SN & 
         $-2.1$ & 
         $-2.3$ \\
         \hline 
         Total & 
         $-8.8$ & 
         $-12.2$ \\
         \hline \hline 
    \end{tabular}
\label{tab:chi2}
\end{table}
We found that non-linear corrections to the matter power spectrum are important when calculating the minimum $\chi^2$.
Specifically, non-linear effects significantly increase the power of the lensing potential power spectrum at small angular scales, which, if not included, worsens the fit to the CMB lensing data.
To ensure a meaningful comparison, we utilize the Halofit prescription~\cite{Bird:2011rb} for the non-linear matter power spectrum for both models.
It is important to note that the Halofit fitting formula was calibrated against $\ld$ and therefore is not expected to accurately capture the non-linear behavior of the matter power spectrum for modified gravity.
Note that the $\wa$ model has two additional parameters compared to $\ld$, whereas the MG model has four extra parameters.

Both models yield nearly the same goodness-of-fit for the DESI and SN data.
The MG scenario shows a better fit to the CMB, but this preference is not statistically significant given the different number of model parameters. 
Overall, the data shows a similar level of preference for both scenarios relative to $\ld$ based on the best-fit $\chi^2$ values:
$2.4\sigma$ for the MG model and $2.5\sigma$ for the $\wa$ model.~\footnote{Ref.~\cite{Wang:2024hks} reported a similar level of preference for modified gravity from the CMB+DESI data when using the $(\mu,\eta)$ parameterization for linear perturbations.}

Additionally, we evaluate the preference for the non-standard scenarios relative to $\Lambda$CDM in the context of a model-selection analysis.
First, we apply the Akaike Information Criteria (AIC)~\cite{1100705}, defined by ${\rm AIC}=\chi^2_{\rm min}+2N_p$, where $N_p$ represents the number of free parameters in the model. 
We found similar preferences for the $\wa$ and MG models relative to $\Lambda$CDM: $\Delta{\rm AIC}=-4.8$ and $-4.2$, respectively.
Next, we employ the Deviance Information Criterion (DIC)~\cite{Liddle:2007fy}, defined as ${\rm DIC}=\chi^2(\theta_p)+2p_D$, where $\theta_p$ represents the parameters at the maximum point of the posterior, and $p_D=\langle\chi^2\rangle-\chi^2(\theta_p)$ is the effective number of parameters that the data constrain, where the average is over the posterior~\cite{Raveri:2018wln}.
We obtain $\Delta{\rm DIC}=-4.9$ and $+1.7$ for the $\wa$ and MG models over $\ld$.
These results suggest a moderate preference for the $\wa$ scenario, while the MG model shows no significant difference from $\ld$, according to the Jeffreys’ scale~\cite{Spiegelhalter:2002yvw}.

In conclusion, while the MG scenario significantly improves the fit quality, it does not show a preference over $\ld$.
The MG model still offers a physical mechanism to safely cross the phantom divide and is capable of accommodate DESI's preference for evolving dark energy.

\section{Conclusions}
\label{sec:conc}

In this work we have explored the preference for a time-evolving dark energy in the light of the DESI DR1 BAO data.
We pinpoint the source of this preference in the $\wa$ model to be due to the the DESI luminous red galaxy samples at $z_\eff=0.51$ and $z_\eff=0.71$, with the latter contributing the most.
We showed that replacing the DESI BAO data at $z<0.8$ with the SDSS measurements in the corresponding redshift range diminishes this preference, making the evolution of dark energy equation of state parameter consistent with $\ld$.
We stress that definitive conclusion from the DESI preference can only be drawn when more data from the DESI DR3 release or from other ongoing and upcoming large surveys will be available.

Taking this preference seriously, we study a physical model which can explain DESI's preference for evolving dark energy in both background and perturbation dynamics.
We explore a general Horndeski scalar-tensor gravity theory, which allows the dark energy equation of state to evolve across the phantom divide, $w=-1$. 
Utilizing the EFT-based framework, which offers a largely model-independent description of modified gravity effects, we derive constraints for $w_0$ and $w_a$ parameters, showing good consistency with the $\wa$ model. 
The overall results show a $2.4\sigma$ discrepancy with $\ld$, as estimated from the best-fit $\chi^2$ values.
We conclude that modified gravity offers a viable physical explanation for DESI's preference for evolving dark energy.

It is important to investigate the robustness of our results with respect to the choice of parameterization of the $\alpha$-functions.
Strictly speaking, the results obtained in this work are valid only for specific subclasses of Horndeski theory where the time evolution of the $\alpha$-functions is proportional to the time evolution of dark energy density in the $w_0w_a$ background.
The primary goal is to extract generic conclusions that are independent of the choice of parametrisation, see e.g.~\cite{Noller:2018wyv}.
We leave a detailed study of parameterization-independent features of the constraints for future work.

Another interesting future avenue is to explore physical modified gravity models based on the full covariant action.
Although this approach offers less generality than the EFT, it allows to directly relate fundamental parameters of the theory to the observational constraints. In the covariant approach, the cosmological background and the linear evolution of dark energy perturbations are determined by the same theory parameters that appear in the action, while in the EFT approach used here, the link between the model parameters and the action is less obvious. This tends to impose a `simplicity constraint' on models specified through a covariant action that
often leads to tighter constraints compared to the EFT-based framework.
However, the most important future development
remains the arrival of new data that may or may not increase the evidence for a departure from $\Lambda$CDM.

\vspace{1cm}
\section*{Acknowledgments}

We are grateful to Emilio Bellini for fruitful discussions. Numerical calculations have been performed with the Baobab high-performance computing cluster at the University of Geneva.

\appendix 

\section{Full Parameter Constraints}
\label{app:full}

\begin{figure*}[ht]
	\begin{center}
		\includegraphics[width=1\textwidth]{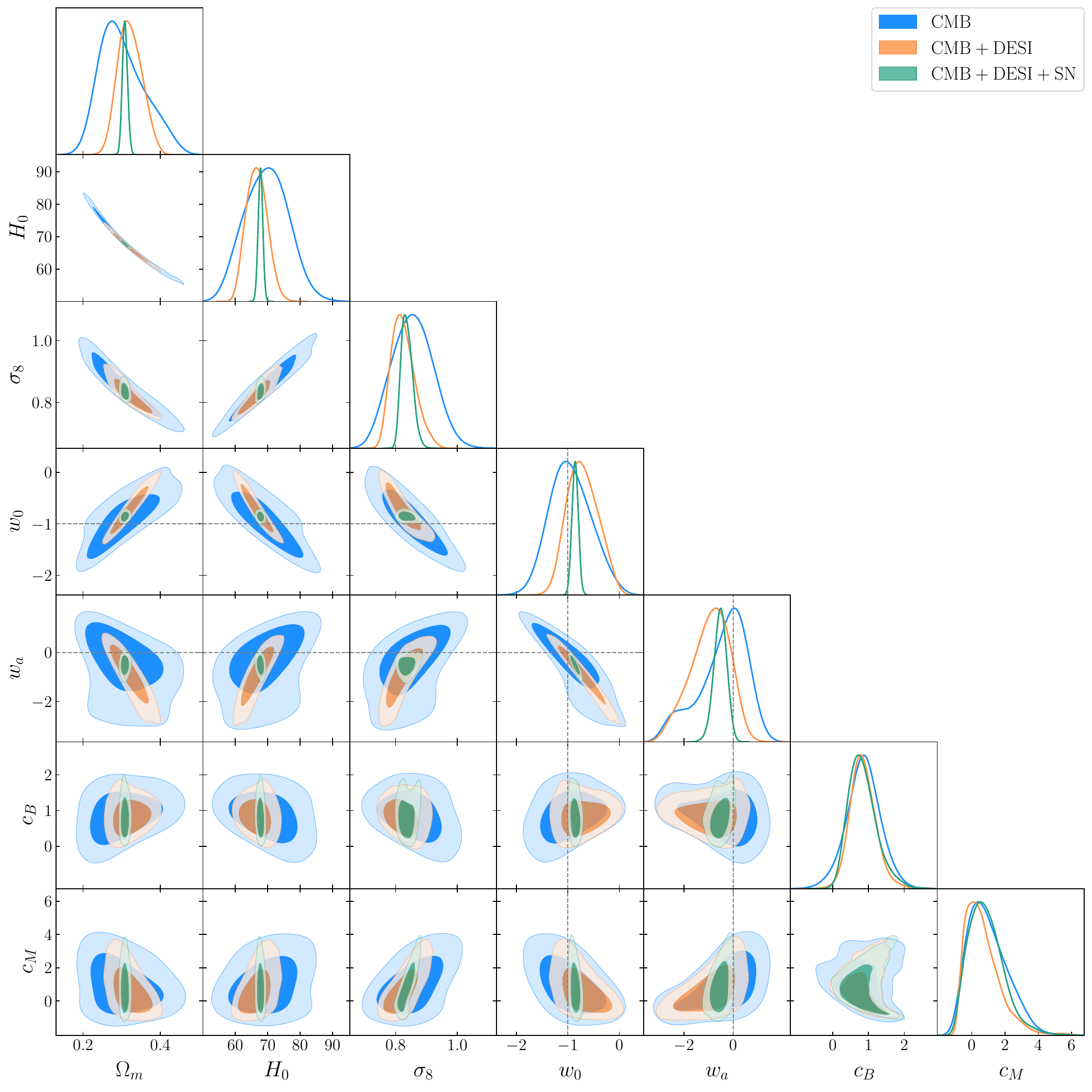}
	\end{center}
	\caption{
		Posterior distributions of the cosmological parameters in the MG model from the CMB (blue), CMB+DESI (orange) and CMB+DESI+SN (green) datasets. 
		\label{fig:MGfull} } 
\end{figure*}

In this appendix, we present the full parameter constraints in the MG model.
Fig.~\ref{fig:MGfull} shows the posterior distributions on cosmological parameters from the CMB, CMB+DESI and CMB+DESI+SN datasets.
The corresponding parameter constraints are listed in Tab.~\ref{tab:par}.
\begin{table*}
\renewcommand{\arraystretch}{1.2}
\caption{Constraints on the relevant cosmological parameters. $H_0$ is measured in units of $\rm km\,s^{-1}\,Mpc^{-1}$.}
\vspace{0.5em}
\centering
\small
\resizebox{\textwidth}{!}{
\begin{tabular}{cccccccc}
\toprule
\hline\hline
Model/Dataset & $H_0$ & $\Omega_{\mathrm{m}}$ & $\sigma_8$ & $w_0$ & $w_a$ & $c_B$ & $c_M$\\
\hline
\textbf{$w_0w_a$CDM} & & & & & &\\
CMB+DESI+SN 
& $67.99_{-0.74}^{+0.72}$ 
& $0.309_{-0.007}^{+0.007}$ 
& $0.818_{-0.010}^{+0.010}$
& $-0.821_{-0.067}^{+0.063}$ 
& $-0.76_{-0.26}^{+0.31}$  
& -- 
& -- \\
\midrule
CMB+(DESI$^\star\!$+SDSS)+SN 
& $67.12_{-0.70}^{+0.68}$ 
& $0.318_{-0.007}^{+0.007}$ 
& $0.811_{-0.010}^{+0.010}$
& $-0.891_{-0.063}^{+0.061}$ 
& $-0.39_{-0.23}^{+0.27}$  
& -- 
& -- \\
\hline
\textbf{MG} & & & & &\\
CMB 
& $68.15_{-6.75}^{+6.63}$ %
& $0.314_{-0.075}^{+0.046}$ %
& $0.854_{-0.070}^{+0.068}$ %
& $-0.857_{-0.447}^{+0.323}$ %
& $-0.42_{-0.61}^{+1.22}$ %
& $0.92_{-0.45}^{+0.42}$ %
& $0.67_{-1.33}^{+0.65}$ %
\\
\midrule
CMB+DESI 
& $66.70_{-3.38}^{+3.16}$ %
& $0.321_{-0.035}^{+0.028}$ %
& $0.825_{-0.044}^{+0.032}$ %
& $-0.721_{-0.355}^{+0.267}$ %
& $-0.91_{-0.60}^{+1.0}$ %
& $0.85_{-0.40}^{+0.34}$ %
& $0.69_{-1.34}^{+0.53}$ %
\\
\midrule
CMB+DESI+SN 
& $67.78_{-0.71}^{+0.73}$ %
& $0.309_{-0.007}^{+0.007}$ %
& $0.837_{-0.022}^{+0.015}$ %
& $-0.859_{-0.062}^{+0.062}$ %
& $-0.52_{-0.25}^{+0.28}$ %
& $0.84_{-0.46}^{+0.31}$ %
& $0.89_{-1.31}^{+0.78}$ %
\\
\midrule
\hline\hline
\end{tabular}}
\label{tab:par}
\end{table*}

Fig.~\ref{fig:MGfull} illustrates the degeneracies between various parameters present in the CMB data. 
The DESI distance measurements significantly improve constraints on the $\ld$ parameters, although the degeneracies among the $w_0$, $w_a$, and other parameters persist.
The addition of SN Ia data breaks the degeneracy in the $w_0-w_a$ plane, thereby refining the estimates of cosmological parameters.

\bibliographystyle{JHEP}
\bibliography{short.bib}

\providecommand{\href}[2]{#2}\begingroup\raggedright\begin{thebibliography}{10}

\bibitem{DESI:2024mwx}
{\scshape DESI} collaboration, A.~G. Adame et~al., \emph{{DESI 2024 VI:
  Cosmological Constraints from the Measurements of Baryon Acoustic
  Oscillations}},  \href{https://arxiv.org/abs/2404.03002}{{\ttfamily
  2404.03002}}.

\bibitem{Scolnic:2021amr}
D.~Scolnic et~al., \emph{{The Pantheon+ Analysis: The Full Data Set and
  Light-curve Release}},
  \href{https://doi.org/10.3847/1538-4357/ac8b7a}{\emph{Astrophys. J.}
  {\bfseries 938} (2022) 113}
  [\href{https://arxiv.org/abs/2112.03863}{{\ttfamily 2112.03863}}].

\bibitem{Rubin:2023ovl}
D.~Rubin et~al., \emph{{Union Through UNITY: Cosmology with 2,000 SNe Using a
  Unified Bayesian Framework}},
  \href{https://arxiv.org/abs/2311.12098}{{\ttfamily 2311.12098}}.

\bibitem{DES:2024tys}
{\scshape DES} collaboration, T.~M.~C. Abbott et~al., \emph{{The Dark Energy
  Survey: Cosmology Results With \textasciitilde{}1500 New High-redshift Type
  Ia Supernovae Using The Full 5-year Dataset}},
  \href{https://arxiv.org/abs/2401.02929}{{\ttfamily 2401.02929}}.

\bibitem{eBOSS:2020yzd}
{\scshape eBOSS} collaboration, S.~Alam et~al., \emph{{Completed SDSS-IV
  extended Baryon Oscillation Spectroscopic Survey: Cosmological implications
  from two decades of spectroscopic surveys at the Apache Point Observatory}},
  \href{https://doi.org/10.1103/PhysRevD.103.083533}{\emph{Phys. Rev. D}
  {\bfseries 103} (2021) 083533}
  [\href{https://arxiv.org/abs/2007.08991}{{\ttfamily 2007.08991}}].

\bibitem{eBOSS:2020mzp}
{\scshape eBOSS} collaboration, A.~J. Ross et~al., \emph{{The Completed SDSS-IV
  extended Baryon Oscillation Spectroscopic Survey: Large-scale structure
  catalogues for cosmological analysis}},
  \href{https://doi.org/10.1093/mnras/staa2416}{\emph{Mon. Not. Roy. Astron.
  Soc.} {\bfseries 498} (2020) 2354}
  [\href{https://arxiv.org/abs/2007.09000}{{\ttfamily 2007.09000}}].

\bibitem{DESI:2024uvr}
{\scshape DESI} collaboration, A.~G. Adame et~al., \emph{{DESI 2024 III: Baryon
  Acoustic Oscillations from Galaxies and Quasars}},
  \href{https://arxiv.org/abs/2404.03000}{{\ttfamily 2404.03000}}.

\bibitem{Wang:2024pui}
Z.~Wang, S.~Lin, Z.~Ding and B.~Hu, \emph{{The role of LRG1 and LRG2's monopole
  in inferring the DESI 2024 BAO cosmology}},
  \href{https://arxiv.org/abs/2405.02168}{{\ttfamily 2405.02168}}.

\bibitem{DESI:2024kob}
{\scshape DESI} collaboration, K.~Lodha et~al., \emph{{DESI 2024: Constraints
  on Physics-Focused Aspects of Dark Energy using DESI DR1 BAO Data}},
  \href{https://arxiv.org/abs/2405.13588}{{\ttfamily 2405.13588}}.

\bibitem{DESI:2024aqx}
{\scshape DESI} collaboration, R.~Calderon et~al., \emph{{DESI 2024:
  Reconstructing Dark Energy using Crossing Statistics with DESI DR1 BAO
  data}},  \href{https://arxiv.org/abs/2405.04216}{{\ttfamily 2405.04216}}.

\bibitem{Yang:2024kdo}
Y.~Yang, X.~Ren, Q.~Wang, Z.~Lu, D.~Zhang, Y.-F. Cai et~al., \emph{{Quintom
  cosmology and modified gravity after DESI 2024}},
  \href{https://arxiv.org/abs/2404.19437}{{\ttfamily 2404.19437}}.

\bibitem{Linder:2007ka}
E.~V. Linder, \emph{{The Mirage of w=-1}},
  \href{https://arxiv.org/abs/0708.0024}{{\ttfamily 0708.0024}}.

\bibitem{Fang:2008sn}
W.~Fang, W.~Hu and A.~Lewis, \emph{{Crossing the Phantom Divide with
  Parameterized Post-Friedmann Dark Energy}},
  \href{https://doi.org/10.1103/PhysRevD.78.087303}{\emph{Phys. Rev. D}
  {\bfseries 78} (2008) 087303}
  [\href{https://arxiv.org/abs/0808.3125}{{\ttfamily 0808.3125}}].

\bibitem{Bellini:2015xja}
E.~Bellini, A.~J. Cuesta, R.~Jimenez and L.~Verde, \emph{{Constraints on
  deviations from \ensuremath{\Lambda}CDM within Horndeski gravity}},
  \href{https://doi.org/10.1088/1475-7516/2016/06/E01}{\emph{JCAP} {\bfseries
  02} (2016) 053} [\href{https://arxiv.org/abs/1509.07816}{{\ttfamily
  1509.07816}}].

\bibitem{Zumalacarregui:2016pph}
M.~Zumalac\'arregui, E.~Bellini, I.~Sawicki, J.~Lesgourgues and P.~G. Ferreira,
  \emph{{hi\_class: Horndeski in the Cosmic Linear Anisotropy Solving System}},
  \href{https://doi.org/10.1088/1475-7516/2017/08/019}{\emph{JCAP} {\bfseries
  08} (2017) 019} [\href{https://arxiv.org/abs/1605.06102}{{\ttfamily
  1605.06102}}].

\bibitem{Frusciante:2018vht}
N.~Frusciante, G.~Papadomanolakis, S.~Peirone and A.~Silvestri, \emph{{The role
  of the tachyonic instability in Horndeski gravity}},
  \href{https://doi.org/10.1088/1475-7516/2019/02/029}{\emph{JCAP} {\bfseries
  02} (2019) 029} [\href{https://arxiv.org/abs/1810.03461}{{\ttfamily
  1810.03461}}].

\bibitem{Bellini:2019syt}
E.~Bellini, I.~Sawicki and M.~Zumalac\'arregui, \emph{{hi\_class: Background
  Evolution, Initial Conditions and Approximation Schemes}},
  \href{https://doi.org/10.1088/1475-7516/2020/02/008}{\emph{JCAP} {\bfseries
  02} (2020) 008} [\href{https://arxiv.org/abs/1909.01828}{{\ttfamily
  1909.01828}}].

\bibitem{Blas:2011rf}
D.~Blas, J.~Lesgourgues and T.~Tram, \emph{{The Cosmic Linear Anisotropy
  Solving System (CLASS) II: Approximation schemes}},
  \href{https://doi.org/10.1088/1475-7516/2011/07/034}{\emph{JCAP} {\bfseries
  1107} (2011) 034} [\href{https://arxiv.org/abs/1104.2933}{{\ttfamily
  1104.2933}}].

\bibitem{Gelman:1992zz}
A.~Gelman and D.~B. Rubin, \emph{{Inference from Iterative Simulation Using
  Multiple Sequences}},
  \href{https://doi.org/10.1214/ss/1177011136}{\emph{Statist. Sci.} {\bfseries
  7} (1992) 457}.

\bibitem{Planck:2018vyg}
{\scshape Planck} collaboration, N.~Aghanim et~al., \emph{{Planck 2018 results.
  VI. Cosmological parameters}},
  \href{https://doi.org/10.1051/0004-6361/201833910}{\emph{Astron. Astrophys.}
  {\bfseries 641} (2020) A6}
  [\href{https://arxiv.org/abs/1807.06209}{{\ttfamily 1807.06209}}].

\bibitem{Carron:2022eyg}
J.~Carron, M.~Mirmelstein and A.~Lewis, \emph{{CMB lensing from Planck
  PR4~maps}}, \href{https://doi.org/10.1088/1475-7516/2022/09/039}{\emph{JCAP}
  {\bfseries 09} (2022) 039}
  [\href{https://arxiv.org/abs/2206.07773}{{\ttfamily 2206.07773}}].

\bibitem{ACT:2023kun}
{\scshape ACT} collaboration, M.~S. Madhavacheril et~al., \emph{{The Atacama
  Cosmology Telescope: DR6 Gravitational Lensing Map and Cosmological
  Parameters}},
  \href{https://doi.org/10.3847/1538-4357/acff5f}{\emph{Astrophys. J.}
  {\bfseries 962} (2024) 113}
  [\href{https://arxiv.org/abs/2304.05203}{{\ttfamily 2304.05203}}].

\bibitem{Planck:2020olo}
{\scshape Planck} collaboration, Y.~Akrami et~al., \emph{{$Planck$ intermediate
  results. LVII. Joint Planck LFI and HFI data processing}},
  \href{https://doi.org/10.1051/0004-6361/202038073}{\emph{Astron. Astrophys.}
  {\bfseries 643} (2020) A42}
  [\href{https://arxiv.org/abs/2007.04997}{{\ttfamily 2007.04997}}].

\bibitem{Ross:2014qpa}
A.~J. Ross, L.~Samushia, C.~Howlett, W.~J. Percival, A.~Burden and M.~Manera,
  \emph{{The clustering of the SDSS DR7 main Galaxy sample \textendash{} I. A 4
  per cent distance measure at $z = 0.15$}},
  \href{https://doi.org/10.1093/mnras/stv154}{\emph{Mon. Not. Roy. Astron.
  Soc.} {\bfseries 449} (2015) 835}
  [\href{https://arxiv.org/abs/1409.3242}{{\ttfamily 1409.3242}}].

\bibitem{BOSS:2016wmc}
{\scshape BOSS} collaboration, S.~Alam et~al., \emph{{The clustering of
  galaxies in the completed SDSS-III Baryon Oscillation Spectroscopic Survey:
  cosmological analysis of the DR12 galaxy sample}},
  \href{https://doi.org/10.1093/mnras/stx721}{\emph{Mon. Not. Roy. Astron.
  Soc.} {\bfseries 470} (2017) 2617}
  [\href{https://arxiv.org/abs/1607.03155}{{\ttfamily 1607.03155}}].

\bibitem{Wang:2024rjd}
D.~Wang, \emph{{The Self-Consistency of DESI Analysis and Comment on ''Does
  DESI 2024 Confirm $\Lambda$CDM?''}},
  \href{https://arxiv.org/abs/2404.13833}{{\ttfamily 2404.13833}}.

\bibitem{Euclid:2024yrr}
{\scshape Euclid} collaboration, Y.~Mellier et~al., \emph{{Euclid. I. Overview
  of the Euclid mission}},  \href{https://arxiv.org/abs/2405.13491}{{\ttfamily
  2405.13491}}.

\bibitem{Bellini:2014fua}
E.~Bellini and I.~Sawicki, \emph{{Maximal freedom at minimum cost: linear
  large-scale structure in general modifications of gravity}},
  \href{https://doi.org/10.1088/1475-7516/2014/07/050}{\emph{JCAP} {\bfseries
  07} (2014) 050} [\href{https://arxiv.org/abs/1404.3713}{{\ttfamily
  1404.3713}}].

\bibitem{Noller:2018wyv}
J.~Noller and A.~Nicola, \emph{{Cosmological parameter constraints for
  Horndeski scalar-tensor gravity}},
  \href{https://doi.org/10.1103/PhysRevD.99.103502}{\emph{Phys. Rev. D}
  {\bfseries 99} (2019) 103502}
  [\href{https://arxiv.org/abs/1811.12928}{{\ttfamily 1811.12928}}].

\bibitem{Wolf:2024stt}
W.~J. Wolf, P.~G. Ferreira and C.~Garc\'\i{}a-Garc\'\i{}a, \emph{{Matching
  current observational constraints with nonminimally coupled dark energy}},
  \href{https://arxiv.org/abs/2409.17019}{{\ttfamily 2409.17019}}.

\bibitem{Bird:2011rb}
S.~Bird, M.~Viel and M.~G. Haehnelt, \emph{{Massive Neutrinos and the
  Non-linear Matter Power Spectrum}},
  \href{https://doi.org/10.1111/j.1365-2966.2011.20222.x}{\emph{Mon. Not. Roy.
  Astron. Soc.} {\bfseries 420} (2012) 2551}
  [\href{https://arxiv.org/abs/1109.4416}{{\ttfamily 1109.4416}}].

\bibitem{Wang:2024hks}
D.~Wang, \emph{{Constraining Cosmological Physics with DESI BAO Observations}},
   \href{https://arxiv.org/abs/2404.06796}{{\ttfamily 2404.06796}}.

\bibitem{1100705}
H.~{Akaike}, \emph{A new look at the statistical model identification},
  \href{https://doi.org/10.1109/TAC.1974.1100705}{\emph{IEEE Transactions on
  Automatic Control} {\bfseries 19} (1974) 716}.

\bibitem{Liddle:2007fy}
A.~R. Liddle, \emph{{Information criteria for astrophysical model selection}},
  \href{https://doi.org/10.1111/j.1745-3933.2007.00306.x}{\emph{Mon. Not. Roy.
  Astron. Soc.} {\bfseries 377} (2007) L74}
  [\href{https://arxiv.org/abs/astro-ph/0701113}{{\ttfamily
  astro-ph/0701113}}].

\bibitem{Raveri:2018wln}
M.~Raveri and W.~Hu, \emph{{Concordance and Discordance in Cosmology}},
  \href{https://doi.org/10.1103/PhysRevD.99.043506}{\emph{Phys. Rev. D}
  {\bfseries 99} (2019) 043506}
  [\href{https://arxiv.org/abs/1806.04649}{{\ttfamily 1806.04649}}].

\bibitem{Spiegelhalter:2002yvw}
D.~J. Spiegelhalter, N.~G. Best, B.~P. Carlin and A.~van~der Linde,
  \emph{{Bayesian measures of model complexity and fit}},
  \href{https://doi.org/10.1111/1467-9868.00353}{\emph{J. Roy. Statist. Soc. B}
  {\bfseries 64} (2002) 583}.

\end{thebibliography}\endgroup

\end{document}